# Role of the Biopolyelectrolyte Persistence Length to Nanoparticle Size Ratio in the Structural Tuning of Electrostatic Complexes

Li Shi,[1,2] Florent Carn,[1] François Boué,[2,3] and Eric Buhler[1,2,*]

[1] *Matière et Systèmes Complexes (MSC) Laboratory, UMR CNRS 7057, University Paris Diderot-Paris 7, Sorbonne Paris Cité, Bâtiment Condorcet, 75205 Paris cedex 13, France*

[2] *Laboratoire Léon Brillouin, UMR 12 CEA-CNRS, CEA Saclay, 91191 Gif-sur-Yvette, France*

[3] *GMPA, UMR INRA 782, 1 avenue Lucien Brétignières, 78850 Thiverval-Grignon, France*

*Corresponding author: eric.buhler@univ-paris-diderot.fr

PACS 62.23.St Complex nanostructures, including patterned or assembled

PACS 82.35.Rs Polyelectrolytes

PACS 83.85.Hf X-ray and neutron scattering

**Abstract.** Aggregation of nanoparticles of given size $R$ induced by addition of a polymer strongly depends on its degree of rigidity. This is shown here on a large variety of silica nanoparticle self-assemblies obtained by electrostatic complexation with carefully selected oppositely charged bio-polyelectrolytes of different rigidity. The effective rigidity is quantified by the total persistence length $L_T$ representing the sum of the intrinsic ($L_p$) and electrostatic ($L_e$) polyelectrolyte persistence length, which depends on the screening, i.e., on ionic strength due to counterions and external salt concentrations. We experimentally show for the first time that the ratio $L_T/R$ is the main tuning parameter that controls the fractal dimension $D_f$ of the nanoparticles self-assemblies, which is determined using small-angle neutron scattering: (i) For $L_T/R<0.3$ (obtained with flexible poly-L-lysine in the presence of an excess of salt), chain flexibility promotes easy wrapping around nanoparticles in excess hence ramified structures with $D_f \sim 2$. (ii) For $0.3<L_T/R\leq 1$ (semiflexible chitosan or hyaluronan complexes), chain stiffness promotes the formation of one-dimensional nanorods (in excess of nanoparticles), in good agreement with computer simulations. (iii) For $L_T/R>1$, $L_e$ is strongly increased due to the absence of salt and repulsions between nanoparticles cannot be compensated by the polyelectrolyte wrapping, which allow a spacing between nanoparticles and the formation of one dimensional pearl necklace complexes. (iv) Finally, electrostatic screening, i.e. ionic strength, turned out to be a reliable way of controlling $D_f$ and the phase diagram behavior. It finely tunes the short-range interparticle potential, resulting in larger fractal dimensions at higher ionic strength.

I. INTRODUCTION

Self-assembly of nanoparticles (NPs) through physical associations offers the possibility of designing bio- and nano-technologically useful supramolecular adaptive nanostructures with precise size, shape and functions [1-7]. The past decade has witnessed great progress in NP self-assembly, yet the quantitative prediction of the architecture of NP ensembles remains a challenge. Hierarchical organizations can be triggered in solution by various mechanisms, such as van der Waals attraction [8], polymerization of functionalized NPs [9,10], lock-and-key binding, depletion [11], magnetic field [12], and electrostatic interactions [13-15] to initiate the self-assembly from the NP level to the mesoscopic one. Among them, the association of charged NPs induced by complexation with (bio)-polyelectrolytes (PELs) of opposite charge is a simple fast, robust, cost-efficient (bio)-process that can lead to new nano-objects, also called electrostatic complexes.

In spite of this large interest, none systematic work has been undergone considering a quite simple aspect as far as chains are concerned: the effect of rigidity of the chain, namely its persistence length, which in a sensible approach, can be compared with the NP size. First approaches have been published: in a preliminary communication [13], we described the complexation between model negatively charged spherical silica NPs with radius $R$~10 nm and chitosan, a natural polyelectrolyte bearing positive charges with a semi-rigid backbone characterized by an intrinsic persistence length of $L_p$~9 nm, using a combination of cryo-TEM, light, small-angle neutron and X-rays scattering. In contrast to strategies based on sophisticated chemical and physical processes (UV photo-reduction [16] or template-directing using membranes or mesoporous media [17], among others), this enabled us to obtain, in an easy and quickly process, well-defined ~250 nm monodisperse nanorods in the presence of an excess of NPs at high ionic strength. Rod-like and fibrous architectures of nano- and mesoscopic dimensions comprise a vast array of nanotechnological and biological machinery but less linear geometries expected to be obtained with more flexible PELs are also of major interest. We, indeed, previously obtained under the same experimental conditions [14] branched complexes of silica NPs instead of nanorods by using a flexible PEL, polylysine.

This result showed the importance of the persistence length to NP radius characteristic ratio. This was, however, a single comparison at high ionic strength; hence the influence of $L_T/R$ was not really established, where $L_T=L_p+L_e$ is the total persistence length accounting for the effective rigidity of the PEL through the sum of two contributions: the intrinsic persistence length $L_p$ of the corresponding uncharged chain and the electrostatic persistence length $L_e$ depending on the ionic strength, which is due to counterions as well as external salt concentration. The aim of the present work is to highlight the pivotal role played by this characteristic ratio $L_T/R$ in the design of nanomaterials in the desired shape and size. For that purpose, we have considered three systems, namely chitosan/NP, hyaluronan/NP and poly-L-lysine/NP complexes, at different PEL and NP concentrations under different electrostatic screening conditions.

Theoretical and experimental studies have actually revealed that along with the physico-chemical parameters such as charge concentration ratio, pH and ionic strength acting on all ionizable species, $L_T$ yields an additional original tuning parameter [13-28]. However, experiments are missing on the influence of NP surface charge density, shape, and size, as well as of the ratio $L_T/R$, the influence of which has been only studied theoretically and by simulations [18-25]. The peculiar case of rigid DNA interacting with nanoparticles was more studied than other PELs due to its biological interest [26,27]. However, the shape of these complexes has been studied only by indirect microscopy methods and not in bulk.

In this article, **we present the structure of the electrostatic complexes for different $L_T$, electrostatic screenings, and polyelectrolytes/NPs pairs so that the influence of the ratio $L_T/R$ can be established**. The NPs we have chosen are spherical silica synthetic NPs that are generally considered as very simple model systems, with fixed characteristics (size, shape, charge, and surface function), regular shape and homogeneous surface. Under well-defined conditions a sensible evaluation of the different kinds of interactions (electrostatic, hydrogen bonds, hydrophobic) can be made showing that electrostatic interactions dominate and can be estimated [13] – in contrast to the more complex case of proteins [28-31].

(i) First, we have studied the complexation between semi-rigid chitosan and ~10 nm radius negatively charged silica nanoparticles (SiNPs). The intrinsic persistence length, $L_p$, of the positively charged chitosan is ~9 nm [32] giving a characteristic ratio $L_T/R$~1 at high ionic strength (in the presence of an excess of salt, the

(i) electrostatic additive contribution to the persistence length $L_e$ is negligible and $L_T \sim L_p$).

(ii) In order to vary this ratio, the polyelectrolyte poly-L-lysine (PLL), displaying positive charges along its flexible backbone, was chosen; the choice of the second partner, 10 nm SiNPs remained unchanged. At high ionic strength (screened conditions), $L_T/R$ is close to 0.1, with $L_p$(PLL)~1 nm [14].

(iii) In a third system, we have reversed the sign of charges of the two components, which means to complex a polyanion with positively charged NPs. In the first place, the NPs are still silica NPs but another surface modification was used to bring positive charges onto their surface. The chosen partner is hyaluronan, a polyelectrolyte polysaccharide displaying a semi-flexible backbone with an intrinsic persistence length $L_p$~5 nm [33-35] that enabled us to get an intermediate ratio, $L_T/R$, close to 0.3 in the presence of an excess of salt. Beyond its well-known role in articulations, hyaluronan is, on a more general basis, found in the extra-cellular matrix, in soft connective tissues as cartilage, where it forms complexes with proteins, glycoproteins and/or other electrostatically charged species. The polymer is assumed to play a role in mitosis and its interaction with extra-cellular polysaccharides has been connected with locomotion and cell migration, which increases its interest in cancer research.

(iv) Finally, the influence of electrostatic interactions was also considered by varying the ionic strength, $I$. In the absence of external salt, the electrostatic contribution to the persistence length $L_e$ depends only on the concentration of the counterions of the two species, allowing a wide variation of $L_T/R$, which is controlled by both the screening and the nature of the PEL.

In what follows, we compare original data from the three systems. The phase behavior in the PEL-NP-$L_T/R$ diagram as well as the structure of the complexes in all characteristic phase domains has been determined using small-angle neutron scattering (SANS) experiments in the presence and absence of external salt. We show that either an increase of $L_T/R$ (equivalent to a reduction of $I$) leads to the formation of lower-dimensional self-assemblies.

## II. EXPERIMENTAL
### A. SAMPLES CHARACTERISTICS

**Chitosan:** Polysaccharide chitosan belongs to a family of linear cationic biopolymers obtained from partial alkaline N-deacetylation of chitin, which is the second most abundant biopolymer on earth. The chitosan studied here is a commercial polymer (with a polydispersity index of around 1.3) from Sigma-Aldrich composed of β 1→4 D-glucosamine units with a degree of N-acetylation equal to 12.5% (determined by NMR) [32]. The mass and the length of the repeating unit are equal to 166 g/mol and 5 Å, respectively. Under acidic conditions, chitosan is water-soluble due to the presence of protonated amino groups. The solutions were then investigated in the presence of 0.3 M acetic acid ($CH_3COOH$) and of 0.2 M sodium acetate ($CH_3COONa$) at high ionic strength. We obtain thus a pH=4.5 buffer where all the amino groups bear a positively charged proton. So chitosan exhibits a high polyelectrolyte character with one positive charge every 5 Å [13,14,32], which would be reduced to one charge per 7 Å after Manning condensation. The intrinsic persistence length of chitosan backbone is roughly equal to 9 nm, due to which it is placed in the class of the so-called semirigid polyelectrolytes. The structural characteristics determined using light scattering, such as the weight average molecular mass ($M_W$=313 000 g/mol), size, and polydispersity are collected in Table 1.

**Poly-L-lysine (PLL):** PLL is a natural flexible homopolymer composed of L-lysine amino acids ( $(C_6H_{12}N_2O)_n$ monomer units have a mass of 128 g/mol) and produced by bacterial fermentation. Each unit of the chain contains an amino group ($NH_3^+$) that renders the whole chain positively charged (pKa=9). The poly-L-lysine hydrobromide used in our study was purchased from Sigma-Aldrich in powder state and was used as received. At high electrostatic screening, aqueous solutions were prepared in the presence of 0.2 M of KBr to keep the same ionic strength than that of chitosan solutions. Under these experimental conditions, all the amino groups are protonated and PLL is fully charged displaying a charge every 3.5 Å (unit size); i.e., every 7 Å after Manning correction. The value of its intrinsic persistence length is roughly equal to $L_p$=1 nm, due to which PLL is classified as flexible PEL [14]. The weight average molecular weight equal to $M_W$=54 000 g/mol (corresponding to 422 charges per chain) was determined in the presence of added salt by static light scattering using a classical Zimm analysis (see Table 1).

**Hyaluronan (HA):** HA or poly((1→3)-β-D-GlcNAc-(1→4)-β-D-GlcA) is a linear semi-flexible polyelectrolyte ($L_p$~5 nm [33-35]) made of a single negatively charged disaccharide repeating unit. We used bacterial hyaluronan produced and carefully purified under the Na salt form by Soliance (Pomacle, France). The monomer mass and length are equal to 401.3 g/mol and 10.2 Å, respectively (global formula $C_{14}H_{20}NO_{11}Na$). At high ionic strength, HA solutions were prepared in the presence of 0.1 M NaCl to ensure the screening of the electrostatic interactions necessary for a chain molecular weight determination using light scattering in dilute regime. A Zimm analysis gives $M_W=9.2\times10^4$ g/mol for the sample referenced "Bashyal" by the supplier (corresponding to 230 charges per chain).

**Preparation of the mixtures:** SiNPs and PELs solutions were prepared separately in either distilled water or in the presence of an excess of salt (0.2 M $CH_3COONa$, 0.2 M KBr, or 0.1 M NaCl for chitosan, PLL, or hyaluronan complexes, respectively, and corresponding to the previously used and standard experimental conditions found in the literature [32-35]) and then diluted and mixed together at various volume ratios to obtain the desired concentrations, ranging over four decades, at constant ionic strength. Mixtures were thoroughly shaken to ensure homogenization and then kept at the temperature of observation, here T=20°C, for several days before visual examination. When a phase separation is observed, samples are re-homogenized and kept at rest for a couple of days to confirm the observations. To ensure a good stability of the positively charged $SiNPs^+$, a buffer at pH=4, obtained by addition of HCl, was used to study the hyaluronan/$SiNP^+$ complexes. Regarding the chitosan/$SiNP^-$ complexes, chitosan is water-soluble only under acidic conditions. The mixtures were then investigated in the presence of 0.3 M acetic acid at pH=4.5 for which both partners are fully charged. Unfortunately, this introduced also a large amount of $CH_3COO^-$ anions in solution, making low ionic strength investigations impossible for this system. According to the supplier, the $SiNPs^-$ have a surface charge density around 1 elementary charge per $nm^2$ giving approximately 1000 negative charges per particle, a value also found by chemical titration [36]. Full characterization of the commercial either negatively or positively charged SiNPs (referenced Ludox AM and CL, respectively by the supplier) are presented in the Appendix (Figs. A1 and A2). Their characteristics, such as NP radii are summarized in Table 1. In all cases, the PEL contour length has been taken larger than the NP circumference in order to ensure the complexation with several NPs.

Table 1. Polyelectrolytes and NPs characteristics. $M_W$, weight average molecular weight determined using static light scattering (SLS); $L_c$, PEL contour length; $R$, SiNPs radius determined using a hard sphere model (see X-rays data analysis in Appendix section); $R_G$, radius of gyration determined either using SLS or small-angle X-ray; $R_H$, hydrodynamic radius determined using dynamic light scattering (DLS); $k_2/k_1^2$, polydispersity index obtained using the cumulant procedure (see Appendix for details); $L_p$, intrinsic persistence length. Error bar is ~10%.

| PELs and NPs | charges | $M_w$ (g/mol) | $L_c$ (nm) | $R$ (nm) | $R_G$ (nm) | $R_H$ (nm) | $k_2/k_1^2$ | $L_p$[b] (nm) |
|---|---|---|---|---|---|---|---|---|
| Chitosan in 0.3M $CH_3COOH$/0.2M $CH_3COONa$ | >0 | 313±20K | 943 | | 66.5 | 44 | 0.24 | 9 |
| Poly-L-lysine in 0.2M KBr | >0 | 54±5K | 148 | | -[a] | 7.9 | 0.2 | 1 |
| Hyaluronan in 0.1M NaCl | <0 | 92±5K | 235 | | -[a] | 14.6 | 0.2 | 5 |
| SiNPs (Ludox AM) in 0.2M $CH_3COOH$ or KBr | <0 | $3 \times 10^6$ | | 9.2 | -[a] | 11.7 | 0.1 | |
| SiNPs (Ludox CL) in 0.1M NaCl | >0 | $3.37 \times 10^7$ | | 17 | 16 | 22.2 | 0.057 | |

[a]Too small to be determined using SLS. [b]At high ionic strength; i.e., in the presence of an excess of salt (0.1 or 0.2 M), the electrostatic additive contribution to the persistence length $L_e$ is negligible and the intrinsic persistence length $L_p$ represents the total persistence length $L_T$.

B. METHODS

**Small-angle neutron scattering (SANS):** SANS experiments were carried out on the PACE spectrometer in the Léon Brillouin Laboratory at Saclay (LLB, France). The chosen incident wavelength, $\lambda$, depends on the set of experiments, as follows. For a given wavelength, the range of the amplitude of the transfer wave vector $q$ was selected by changing the detector distance, $D$. For poly-L-lysine samples, three sets of sample-to-detector distances and wavelengths were chosen ($D = 1$ m, $\lambda = 6 \pm 0.5$ Å; $D = 4.7$ m, $\lambda = 6 \pm 0.5$ Å and $D = 4.7$ m, $\lambda$

= 13± 0.5 Å) so that the following $q$-ranges were respectively available: $3.63 \times 10^{-2} \leq q$ (Å$^{-1}$) $\leq 3.7 \times 10^{-1}$, $6.88 \times 10^{-3} \leq q$ (Å$^{-1}$) $\leq 7.33 \times 10^{-2}$, and $3.18 \times 10^{-3} \leq q$ (Å$^{-1}$) $\leq 3.38 \times 10^{-2}$. In a second run (hyaluronan complexes) we used $D = 1$ m, $\lambda = 6 \pm 0.5$ Å; $D = 3$ m, $\lambda = 12 \pm 1$ Å, and $D = 4.5$ m, $\lambda = 17 \pm 0.5$ Å so that the following $q$-ranges were respectively available: $3.63 \times 10^{-2} \leq q$ (Å$^{-1}$) $\leq 3.7 \times 10^{-1}$, $5.13 \times 10^{-3} \leq q$ (Å$^{-1}$) $\leq 5.45 \times 10^{-2}$, and $2.43 \times 10^{-3} \leq q$ (Å$^{-1}$) $\leq 2.59 \times 10^{-2}$. Finally, in a third run (chitosan samples) we used: $D = 1$ m, $\lambda = 10 \pm 1$ Å and $D = 4.7$ m, $\lambda = 10 \pm 1$ Å so that the following $q$-ranges were respectively available: $2.2 \times 10^{-2} \leq q$ (Å$^{-1}$) $\leq 2.1 \times 10^{-1}$ and $4.15 \times 10^{-3} \leq q$ (Å$^{-1}$) $\leq 4.4 \times 10^{-2}$. Measured intensities were calibrated to absolute values (cm$^{-1}$) using normalization by the attenuated direct beam classical method. Standard procedures to correct the data for the transmission, detector efficiency, and backgrounds (solvent, empty cell, electronic, and neutronic background) were carried out. The scattering wave vector, $q$, is defined by eq 1, where $\theta$ is the scattering angle:

$$q = \frac{4\pi}{\lambda} \sin\frac{\theta}{2}. \qquad (1)$$

The usual equation for absolute neutron scattering combines the intraparticle scattering $S_1(q) = V\phi P(q)$ factor ($P(q)$ is the form factor) with the interparticle scattering $S_2(q)$ factor

$$I(q)(cm^{-1}) = (\Delta\rho)^2 (S_1(q) + S_2(q)) = (\Delta\rho)^2 (V\phi P(q) + S_2(q)), \qquad (2)$$

where $(\Delta\rho)^2 = (\rho_{solute} - \rho_{solvent})^2$ is a contrast per unit volume between the solute and the solvent and was determined from the known chemical composition. $\rho = \Sigma n_i b_i / (\Sigma n_i m_i v \times 1.66 \times 10^{-24})$ represents the scattering length per unit volume, $b_i$ is the neutron scattering length of the species i, $m_i$ the mass of species i, and $v$ the specific molecular volume of the solute (PEL monomer and/or nanoparticle - see Table of the Appendix for contrast and specific volume values) or the solvent (i.e., 0.91 cm$^3$g$^{-1}$ for D$_2$O). $P(q)$ is the form factor of the scattered objects, $V = Nvm \times 1.66 \times 10^{-24}$ is the volume of the N sub-units (of mass $m$) comprising the scattered objects and $\phi$ is their volume fraction. In the high $q$-range, the scattering is assumed to arise from isolated scattered objects; i.e., $S_2(q) = 0$, and thus $I(q) \propto P(q)$.

Because of the high molecular weight of the SiNPs and concentrations used in this study, the SANS signal of the complexes mostly arises from the scattering of the nanoparticles. Consequently, the scattered objects may be described as complexes made of several NPs linked to each other by polyelectrolyte chains of negligible scattering.

All SANS experiments were performed in $D_2O$.

**Dynamic and static light scattering experiments:** The measurements used a 3D DLS spectrometer (LS Instruments, Fribourg, Switzerland) equipped with a 25mW HeNe laser (JDS uniphase) operating at $\lambda=632.8$ nm, a two channel multiple tau correlator (1088 channels in autocorrelation), a variable-angle detection system, and a temperature-controlled index matching vat (LS Instruments). The scattering spectrum was measured using two single mode fibre detections and two high sensitivity APD detectors (Perkin Elmer, model SPCM-AQR-13-FC). Solutions were filtered through 0.2 μm PTFE Millipore filter into the cylindrical scattering cell. Methods used to characterize PELs and NPs using light scattering are detailed in the Appendix.

## III. RESULTS
### A. INFLUENCE OF THE RATIO $L_T/R$ IN THE PRESENCE OF EXTERNAL SALT

**Phase behavior:** First the phase evolution at high ionic strength (electrostatic screened conditions with a Debye length $\kappa^{-1}\leq 1$ nm obtained by addition of a large excess of salt) shows notable similarities between the three investigated PEL/SiNP pairs (see Fig. 1). Three characteristic domains are visualized in the PEL-SiNP concentration plane of the phase diagram: mixtures are monophasic and transparent in the presence either of an excess of PELs (hereinafter called domain I) or of NPs (domain III). A biphasic region (domain II) is observed for intermediate concentration ranges, where one rich and turbid phase of coacervate coexists with an upper dilute and limpid phase (defined as supernatant). The volume ratio of the two phases, separated by a net and sharp interface, depends on the concentration ratio. Boundaries between domains may be slightly shifted in deuterated water.

Such sequence of phase behaviors is reminiscent of "complex coacervation" due to electrostatic attraction between PELs and oppositely charged NPs [14,28-31]. The two partners can first associate into primary complexes, which are neutral. Above a certain concentration they precipitate in rich and poor liquid phases and form fractal aggregates by controlled (diffusion limited or reaction limited) aggregation, or solid clusters depending on the potential profile. Another aspect is the release of counter-ions of both species, which has an entropic contribution on the free energy [30,37]. The stoichiometric threshold corresponding to a charge ratio [+]/[-]=1 cuts the biphasic region in half, which agrees with the electrostatic origin of complexation (see continuous lines in Fig. 1). Here we have assumed that all counter-ions of both species are released and have ignored Manning condensation, if applicable (e.g. for chitosan and PLL). In practice, SANS results indicate that the dense lower phase contains the major part of the NPs and PELs, while the upper phase is a very dilute solution of complexes. It is noteworthy that the phase separation is observed for minute quantities of PEL.

Interestingly, the boundary between domains II and III is shifted downwards for flexible PLL complexes, whereas that between domains I and II seems to remain unchanged within the error bars. In other words, region II stretches towards larger NP contents when one diminishes the $L_T/R$ ratio, and hence when one may change the inner structure and compactness of complexes. It should be remembered here that the electrostatic additive contribution to the persistence length $L_e$ is negligible in the presence of 0.1 or 0.2 M of excess salt (Debye screening length $\kappa^{-1}$<1 nm). As a consequence, the intrinsic persistence length $L_p$ represents the total persistence length at high ionic strength.

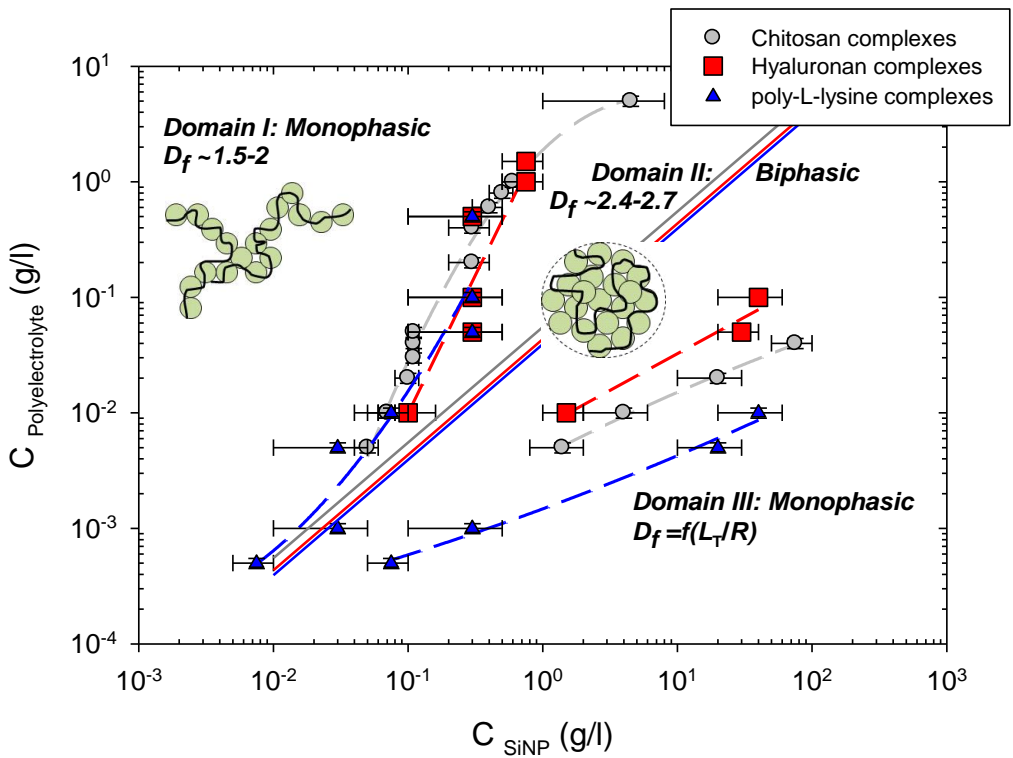

Fig. 1. (Color on-line) Sequence of phase behaviors in the PEL-SiNP concentration plane at T=20°C and high ionic strength. For chitosan, hyaluronan, and PLL systems, mixtures are prepared in H$_2$O in the presence of 0.2M CH$_3$COONa, 0.1M NaCl, and 0.2M KBr, respectively. The continuous grey, red, and blue continuous lines indicate the stoichiometric thresholds corresponding to a charge ratio [+]/[-]=1 for chitosan, HA, and PLL complexes, respectively.

**Structure of the complexes:** To gain insight into the structure of the complexes, SANS experiments were performed at LLB (beamline PACE) on mixtures prepared in D$_2$O and representing the different parts of the phase diagram. Because the signal is dominated by the SiNPs, especially in domains II and III (as checked by contrast matching experiments cancelling the polymer signal, not shown), SANS is an appropriate method for determining the NPs arrangement over the range of 1-30 nm. In the experimental scattering vector regime $q>R_g^{-1}$ (where $R_g$ is the radius of gyration of the objects), the scattered intensity of a fractal particle is given as $I(q)\sim q^{-D_f}$. Here $D_f$ is the so-called fractal dimension of the particle, which determines the scaling of the mass of the clusters with its size $M\sim R_g^{D_f}$. If log$I(q)$ is plotted vs. log$q$, one obtains a linear decay with slope $D_f$, allowing us to directly determine the fractal dimension of the complexes, subject to the condition of a linear decay over at least one order of magnitude of the experimental $q$-regime. This is what is observed in most cases: a low-$q$ Guinier regime associated to the finite size of the complexes is rarely visible, indicating that NP self-assemblies are larger than 30 nm. Among the very many scattering profiles, eighteen of them are visualized in Figs 2, 3, and 4. For complexes in domain I (excess of PELs), patterns characteristic of ramified or branched structures with $D_f$ being between 1.5 (hyaluronan complexes) and 2 (chitosan and PLL complexes) are observed at high ionic strength (see Fig. 2). For the biphasic domain II, the scattering varies as a power law with exponent $D_f$ ranging from 2.4 and 2.7 depending on the concentration ratio, characteristic of rather compact fractal aggregates (Fig. 3) [38]. This behavior, already reported on several systems such as protein/PEL [29-31] or NP/polymer complexes [14], among others, is observed in the whole biphasic domain for coacervates as well as for supernatant phases as long as their scattering level is not negligible suggesting that complexes are formed before separation.

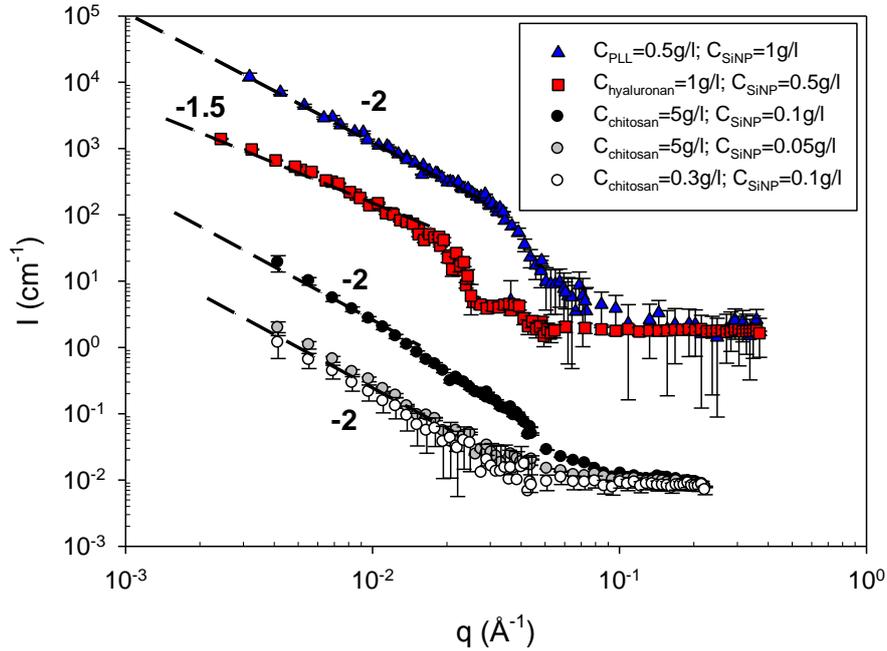

Fig. 2. (Color on-line) SANS profiles collected in domain I (excess of PELs) at high ionic strength (presence of an excess of salt) and 20 °C of the three systems examined. For clarity, the HA and the PLL spectra have been shifted by two and three log units along the y-axis, respectively. While the 0.5 g/l PLL/1 g/l SiNP sample is biphasic in $H_2O$ (close to the separation line between domains I and II), it is monophasic and representative of domain I in $D_2O$, showing the slight isotopic effect on the position of the phase transition line.

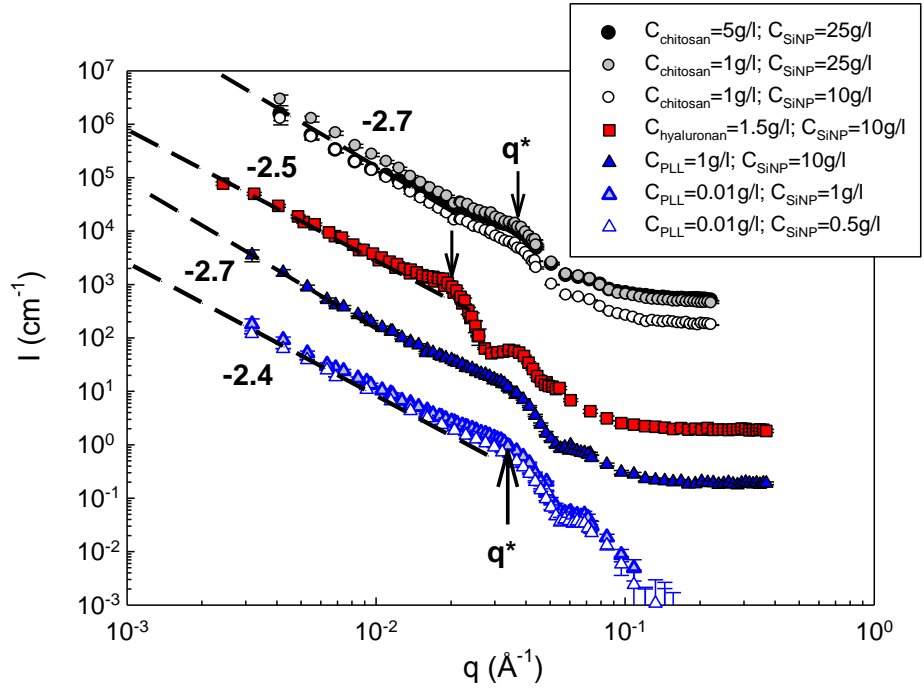

Fig.3. (Color on-line) SANS profiles collected in domain II (coacervates) at high ionic strength (presence of an excess of salt) and 20 °C of the three systems examined. For clarity, spectra have been shifted by one or two log units along the y-axis with respect to each other. The arrows indicate the position q* of a characteristic correlation peak (see text, part III.B).

The most striking result is obtained in domain III (excess of NPs), where the behavior is profoundly different from one system to another one and depends on $L_T/R$. Figs. 4a and 4b show representative SANS profiles for each of the three systems in the presence of an excess of SiNPs. The three spectra are markedly different, showing that morphologically different structures are present in solution. For flexible poly-L-lysine complexes, the $q^{-2}$ dependence suggests a moderately ramified distribution for the SiNPs inside the complexes or branched aggregates. This result profoundly differs from that obtained with semiflexible chitosan ($L_T/R=1$) showing a $q^{-1}$ law due to the formation of nanorods under the same experimental conditions. Here PLL, a flexible polyelectrolyte, with $L_T/R\sim0.1$ cannot induce the formation of 1D structures. Semiflexible hyaluronan with $L_T/R=0.3$ give rise to intermediate topology complexes with $D_f=1.5$. A power law fit (shown as a guide to the eyes in inset of Fig. 4a) gives $D_f\sim(L_T/R)^{-0.3}$ and shows the major role played by the polyelectrolyte persistence length on the compactness of the NP complexes. Comparing with theory, we see that these trends are consistent with Monte Carlo simulations that have been used to examine the complexation of linear PEL possessing variable flexibility with several oppositely charged macroions. In particular, Jonsson and Linse [24,25] found that in excess of macroions (corresponding here to domain III) the polyelectrolyte complexe becomes consistently overcharged due to charge reversal and that the macroions (or NPs) might come into molecular contact with each other despite their mutual repulsion. When the chain is flexible or nearly flexible the complexed macroions form a compact structure, which opens up as the chain becomes stiffer. For the most rigid PEL chain, the macroions become nearly linearly arranged with fewer polyelectrolyte segments near the NP surface. At even larger amount of NPs, there exist uncomplexed NPs separated from the complexes, the number of which has also been determined experimentally [13,14].

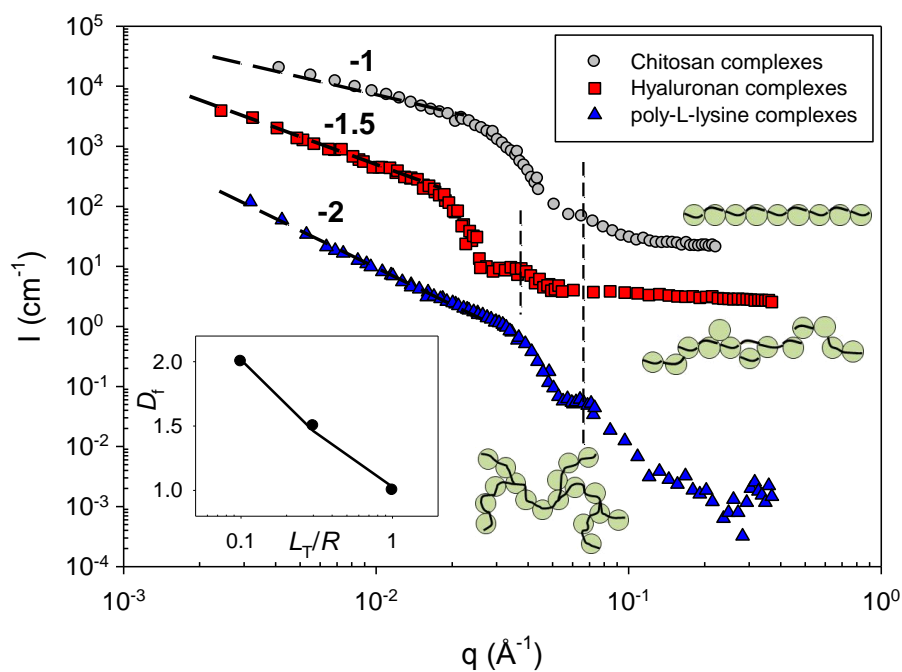

Fig. 4a. (Color on-line) SANS profiles collected at high ionic strength, $I$, and 20°C of the three PEL complexes: 0.01g/l chitosan/10g/l SiNP, 0.01g/l HA/2g/l SiNP, and 0.01g/l PLL/10g/l SiNP (monophasic and representative of domain III in $D_2O$) solutions are prepared in the presence of 0.2M $CH_3COONa$, 0.1M NaCl, and 0.2M KBr, respectively. For clarity, hyaluronan and chitosan complexes spectra have been shifted by two and three log units along the y-axis with respect to that of PLL complexes, respectively. All curves exhibit the first oscillation associated to the form factor of the SiNPs cross-section occurring around $6.5 \times 10^{-2}$

Å$^{-1}$ (9.2 nm SiNP$^-$) or 3.5×10$^{-2}$ Å$^{-1}$ (17 nm SiNP$^+$) for chitosan and PLL or hyaluronan complexes, respectively (see dashed lines and appendix). The inset represents the variation of $D_f$ with the characteristic ratio $L_T/R$ (in the presence of 0.1 or 0.2 M external salt, $L_T=L_p$).

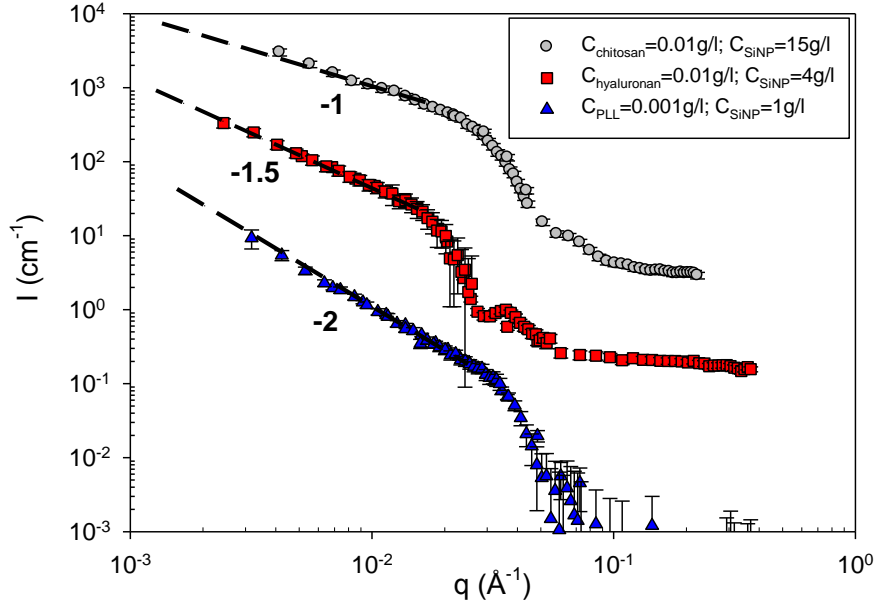

Fig. 4b. Same as (a), but for other PEL and NP concentrations (domain III, excess of NPs). For clarity, the hyaluronan data have been shifted by one log unit along the y-axis with respect to those of PLL.

Interestingly, as in domains I and II (Figs. 2 and 3, respectively), experiments performed at different $C_{PEL}/C_{NP}$ ratios in domain III of the phase diagram show the same structure for the complexes (same $D_f$).

B. SALT-FREE MIXTURES AND INFLUENCE OF IONIC STRENGTH

Ionic strength, $I$, is proving to be also a reliable means of controlling the ratio $L_T/R$ and compactness of NPs self-assemblies. We will see that it controls as well the extent of the phase diagram regions. In the absence of external salt, we find the same sequence of phase behaviors as that reported in the former section: monophasic domains in the presence of either a PEL or a NP excess, and a two-phase region for intermediate NP contents. An example is provided in Fig. 5, which shows these domains for hyaluronan complexes with and without addition of salt. The main difference comes from the boundary between domains II and III, which is shifted to lower SiNP concentrations at low *I*. Obviously charges screening reduces electrostatic repulsion in favour of van der Waals attraction, hence an easier complexation causing the extent of the coacervate region at high *I*. Such behavior is independent on the nature of the PEL and on the sign of charge.

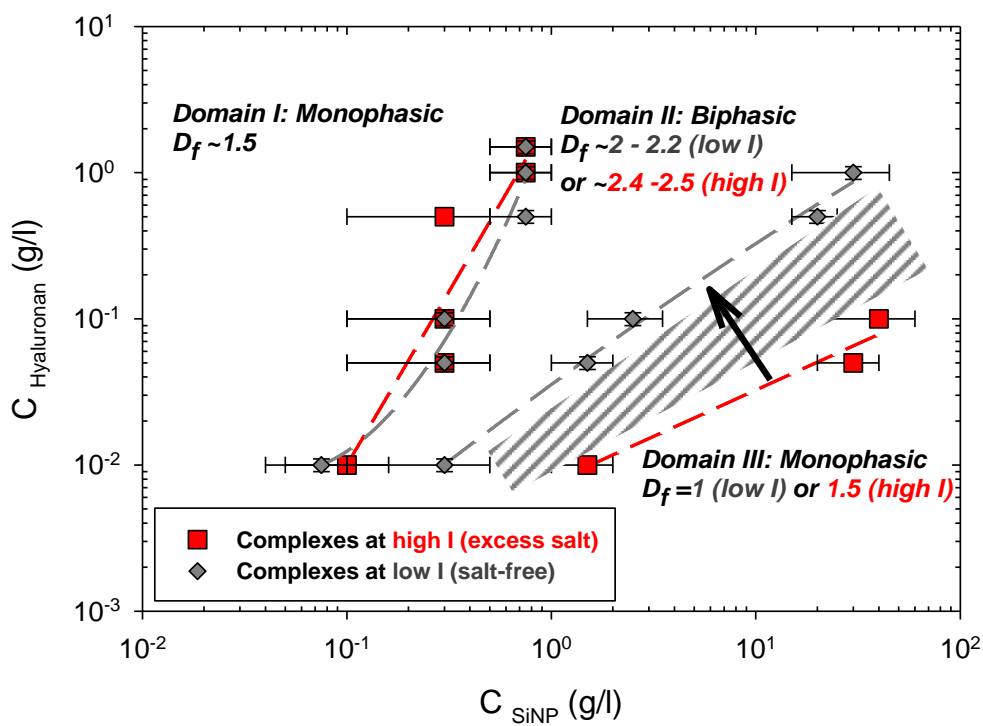

Fig. 5. (Color on-line) Sequence of phase behaviors in the HA-SiNP concentration plane at high (presence of external salt) and low (salt-free systems) ionic strength, *I*. The red dashed lines represent boundaries between characteristic domains at high *I*, whereas gray dashed lines represent boundaries at low *I*. The shaded area corresponds to the coacervate region expansion when *I* is increased (diagram obtained in $H_2O$).

The effect of salt addition on the phase diagram, showing the importance of electrostatic screening is confirmed by SANS: more compact structures are obtained in coacervates as well as in monophasic domains at high ionic strength as shown by the larger fractal exponents, $D_f$, determined in the presence of an excess of salt. SANS spectra obtained for all investigated systems show the same trend and can be visualized in Figs. 6 and 7: Lower $D_f$ are obtained in the absence of external salt in all domains of the phase diagram (in salt-free mixtures the Debye screening length depends only on the counterions concentrations and is ranging between 5 and 23 nm depending on the samples). For clarity, the many results are also summarized in Table 2. Other examples are given in Fig. 8, which shows the scattering profiles collected for representative HA samples in domains II and III in the presence and absence of excess salt.

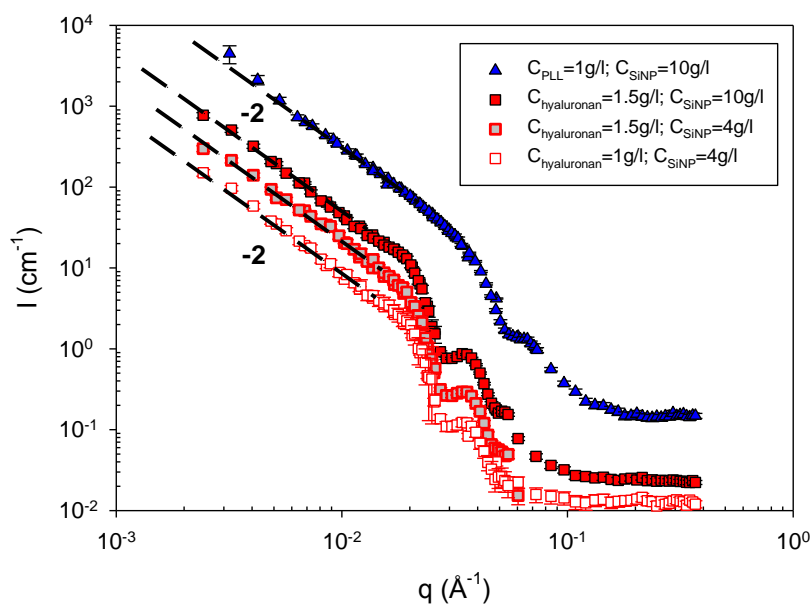

Fig. 6. (Color on-line) SANS patterns collected in domain II (coacervates) at low ionic strength (salt-free mixtures) and 20 °C of the two systems examined. For clarity, the PLL data have been shifted by one log unit along the y-axis with respect to those of HA.

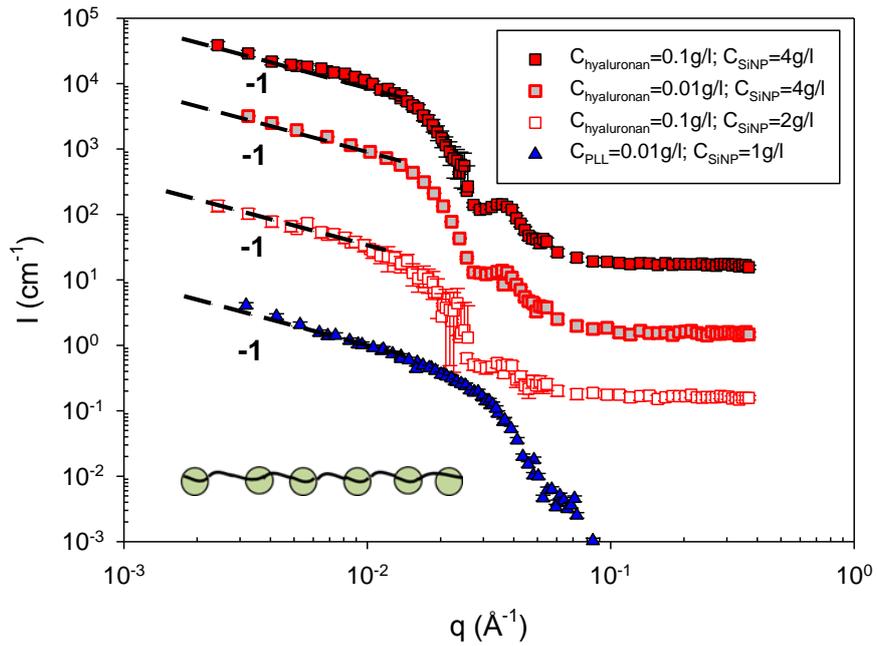

Fig. 7. (Color on-line) SANS patterns collected in domain III (excess of NPs) at low ionic strength (salt-free mixtures) and 20 °C of the two systems examined. For clarity, the spectra have been shifted by one log unit along the y-axis with respect to each other. While the $C_{hyaluronan}$=0.1 g/l; $C_{SiNP}$=2 g/l sample is biphasic in $H_2O$ (close to the separation line between domains II and III), it is monophasic and representative of domain III in $D_2O$, showing the slight isotopic effect on the position of the phase transition line.

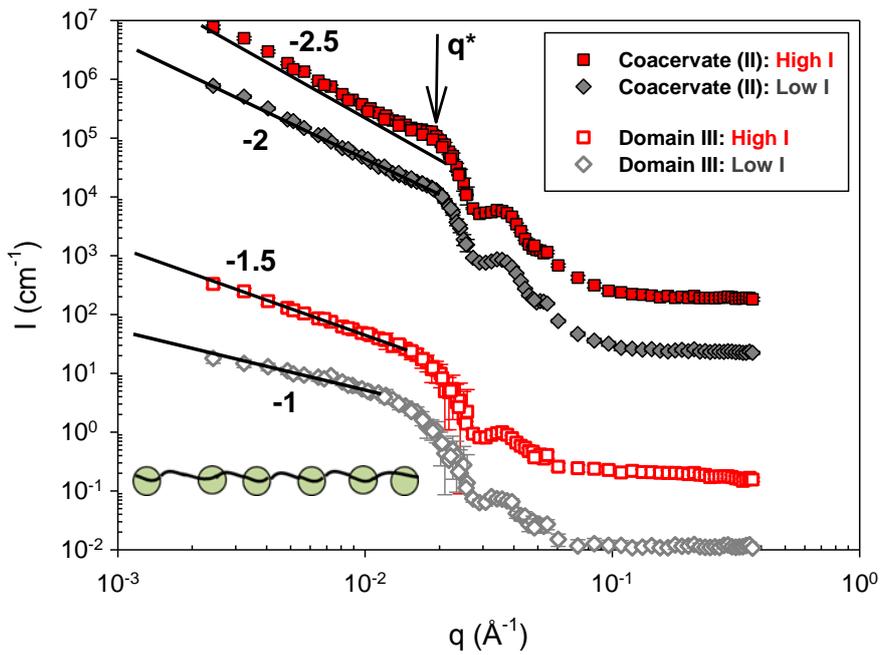

Fig. 8. (Color on-line) SANS profiles collected at high (0.1 M NaCl) and low ionic strength (salt-free mixtures) for HA complexes in domains II and III. Concentrations are $C_{HA}$=1.5 g/l and $C_{SiNP}$=10 g/l for coacervates (filled symbols), whereas concentrations are equal to $C_{SiNP}$=2 g/l and $C_{HA}$=0.01g/l (at high I) or 0.05 g/l (at low *I*) for mixtures in domain III (open symbols). For clarity, the spectra have been shifted by one or two log units along the y-axis with respect to each other. *q\** indicates the position of the characteristic correlation peak.

Also, the phase separation kinetics is always much faster in the presence of an excess of salt. The opaque coacervate falls down instantaneously and is very viscous, even gel-like, whereas the supernatant is very clear, fluid, and diluted [39]. This increase of separation rate is accompanied by larger fractal exponent values for coacervates as is for example the case for HA complexes for which $D_f$~2.5, a value definitively larger than 2 determined for salt-free mixtures. Coacervates are also more viscous at high *I*. Additionally, we observe the emergence of a correlation peak, which has an abscissa $q^*$=0.018 Å$^{-1}$, corresponding to the distance, $d=2\pi/q^*$=350 Å~$2R_{SiNP+}$, between positively charged SiNPs in close contact (HA complexes, see Figs. 3 and 8). For chitosan and PLL coacervates, this peak is observed at 0.035 Å$^{-1}$ and matches with the characteristic distance $d$=180 Å~$2R_{SiNP-}$ between negatively charged silica nanoparticles in close contact (see Fig. 3).

Table 2. Fractal exponents, $D_f$, obtained in the low-q regime (considered samples are shown in Figs. 2, 3, 4, 6, 7, and 8). $L_T$ is the PEL total persistence length and $R$ the real NPs radius determined by fitting the data by means of the form factor expression derived for hard spheres (see appendix).

|  | Chitosan/SiNP⁻ | | PLL/SiNP⁻ | | HA/SiNP⁺ | |
|---|---|---|---|---|---|---|
|  | $L_T/R$ | $D_f$ | $L_T/R$ | $D_f$ | $L_T/R$ | $D_f$ |
| *Domain (I):* Excess salt | 1 | 2 (ramified) | 0.1 | 2 | 0.3 | 1.5 |
| Salt-free | - | - | 3 | $\geq 1$ [a] | 1 | $\geq 1$ [a] |
| *Coacervate (II):* Excess salt | 1 | 2.7 (compact) | 0.1 | 2.4-2.7 | 0.3 | $\geq 2.4$ |
| Salt-free | - | - | 1 | 2 | $1 \geq L_T/R \geq 0.7$ | 2 |
| *Domain (III):* Excess salt | 1 | 1 (nanorods) | 0.1 | 2 | 0.3 | 1.5 |
| Salt-free | - | - | 19 | 1 (pearl necklaces) | $6 \geq L_T/R \geq 3$ | 1 (pearl necklaces) |

[a] Too poor statistics due to low NPs and PELs concentration.

## IV. DISCUSSION

### A. EFFECT OF *I* ON THE STRUCTURE OF THE COMPLEXES: AN EFFECT OF THE PERSISTENCE LENGTH?

Two effects are quite clear from results above: the one of persistence length, and the one of ionic strength. We want now to discuss more in detail how these two quantities can be combined in a single one, the total persistence length $L_T$.

**The ingredients for the total persistence length**: we first need to discuss the values for the persistence length. When electrostatic repulsions between the charges along the chain are not screened, they will tend to make even larger the local rigidity and increase the global size of the polyelectrolyte. An additional electrostatic persistence length, $L_e$, due to electrostatic repulsions increases the effective persistence length, $L_p$. Then, following the simplest model, the total persistence length represents the sum of these two contributions: the intrinsic persistence length $L_p$ of the corresponding uncharged chain and the electrostatic persistence length $L_e$, which depends on the screening, i.e., on counterions amount as well as on external salt concentration [40]:

$$L_T = L_p + L_e, \text{ with } \kappa L_T \gg 1 \qquad (3)$$

This model was first proposed for polyelectrolytes near the rod limit, hence for $L_p$ large enough, but extended to the case where $L_p+L_e$ is large enough. Assuming a Debye-Hückel potential and under the condition that counterions condensation occurs when necessary, Odijk [40] and Skolnick and Fixman [41] found:

$$L_e = \frac{\xi^2}{4\kappa^2 l_B} \text{ for } \xi < 1, \text{ and } L_e = \frac{1}{4\kappa^2 l_B} \text{ for } \xi > 1 \qquad (4)$$

where $l_B$=7.13 Å in water is the Bjerrum length, $\kappa^{-1}$ is the Debye-Hückel screening length related to the concentration of the counterions, and $\xi=l_B/a$ is the structural charge parameter, where $a$ is the distance between two ionic sites. For $\xi>1$, Manning's counterions condensation is expected to bring the distance between the charges along the chain down to $l_B$. In dilute solutions $\kappa^2=4\pi l_B c_f$, where $c_f$ is the concentration of free monovalent ions. If $\xi<1$, $c_f=c+2c_s$, where $c$ is the concentration of PELs and SiNPs counterions and $c_s$ the excess salt concentration. If $\xi>1$, part of the PEL counterions are condensed and $c_f=c/\xi +2c_s$.

For hyaluronan (HA), Manning's counterions condensation is not expected, since the distance between two ionic sites $a$=10.2 Å is larger than $l_B$ ($\xi$=0.7 <1). Also, $L_p$ = 5 nm, which does not correspond to the rigid rod limit, but agrees with eqs. 3 and 4, since absolute values of $L_e$ and the variation $L_e \sim c_f^{-1}$ have been validated previously on HA solutions [34,35]. In domain III, the corresponding values for $L_e$, which depend on the concentration of both partners, range between 45 and 175 nm ($\kappa^{-1}$ is ranging between ~16 and 23 nm), hence the condition $\kappa.L_T >1$ is fulfilled.

For chitosan and PLL, $\xi>1$ and Manning's condensation brings the distance between charges, $a$, down to $l_B$. Although, we do not have data for chitosan at low $I$, we have data for PLL. The value $L_p$=1 nm corresponds to a flexible PEL, but we will still apply eq. 4: indeed the condition $\kappa.L_T >1$ is still fulfilled because $L_T$ is large enough due to the contribution of $L_e$.

Finally, for salt-free mixtures (low $I$) in domain III, due to the contribution of $L_e$, the ratio $L_T/R$ could be as large as ~6 for hyaluronan ($R$=17 nm for positively charged Ludox CL), and

~19 for PLL complexes (since $R$ is smaller, 9.2 nm for negatively charged Ludox AM). At low $I$, it decreases to a value comprised between 1 and 3 in domain I, and between 0.7 and 1 in domain II ($L_e$ ranging between 6.6 and 11 nm with $5 \leq \kappa^{-1} \leq 8$ nm).

At larger $I$, i.e. with the quantities of added salt used here, $\kappa^{-1} < 1$ nm, so that $L_e$ becomes negligible and $L_T \sim L_p$.

**Elongated complexes in domain III.** Ionic strength $I$ has a strong influence: while HA and PLL complexes form elongated ($D_f$=1.5) or ramified objects ($D_f$=2) in domain III at large $I$ (0.1 or 0.2 M added salt), they adopt a rodlike structure with $D_f$=1 at low $I$ (Figs. 7 and 8). The $q^{-1}$ variation is then fitted satisfactorily by means of the form factor derived for rigid rod particles, $\pi/qL$, where $L$ is the length of the rod, [14] in order to determine the linear mass density of the aligned SiNPs in the HA complexes, $M_L$. Considering that the scattered intensity, $I=(\Delta\rho)^2 \times V \times \phi \times P(q)$, arises only from the NPs – the signal of PELs is negligible here – we obtain $M_L=(I \times q)/(1.66 \times 10^{-24} \times (\Delta\rho)^2 \times v \times \phi \times \pi)$=19305 g/mol/Å for all HA and SiNP concentrations in domain III (see Figs. 7 and 8), a value five times lower than that calculated for straight monolayer wires of NPs in close contact (given by the NPs mass to their diameter ratio equal to $3.38 \times 10^7/340$=99411 g/mol/Å). This shows that, in the absence of excess salt, these NP complexes adopt a rodlike structure with some non-covered chain parts, which we can call "holes" or also pearl necklaces (see drawing in Fig. 8). An explanation for these "holes" comes from the increase of the HA chain stiffness at low $I$ due to electrostatic repulsion between monomers that gives rise to a more stretched chain with fewer PEL segments near the NP surface. It has been previously proposed by theorists [22,24] that if the charge of a NP is not completely compensated by the PEL wrapping, the net charge of the NP is still positive, and the neighboring NPs repeal each other and can form nanorods with holes. Although NPs are not in close contact on average, stretched PELs ensure their linking within the rodlike assemblies.

**$L_T$: a unique parameter.** In the sake of generality, we can check whether the influence of $I$ is accounted by a general dependence over a unique parameter, the total persistence length $L_T$. In Fig. 9, we have regrouped all $D_f$ values for the different PELs and the different ionic strengths for domain III. We observe a simple behavior. First $D_f$ =1 for $L_T/R \geq 1$; this comprises low $I$ data (right hand side), but also the (more rigid) chitosan at large $I$ for which $L_T/R$ is just equal to 1. Second, for $L_T/R <1$, $D_f$ values for HA as well as PLL increase up to 2. Hence a general, clear and systematic account of $I$ through $L_T$, is obtained here. We detail it just below.

For $L_T/R \leq 1$, the PEL wrapping compensates the charge of the NP that might come in molecular contact with each other. Therefore, the complexed NPs form a compact structure, which opens up as the chain becomes stiffer. For $L_T/R=1$, the NPs become linearly arranged with fewer segments near the NP surface. For $L_T/R>1$, repulsions between NPs along the rigid chain cannot thus be compensated by the wrapping of the PEL, which enables a spacing between NPs and the formation of pearl necklace-like structures with $D_f=1$. This is obtained for HA/SiNP complexation, but such effect also accounts for the formation of elongated PLL complexes with $D_f \sim 1$ at low $I$ in the same domain III, although this PEL on its own displayed a very flexible backbone. Interestingly, the linear mass density of the PLL rodlike complexes, $M_L=3500$ g/mol/Å, is also about five times lower than that of SiNPs aligned in close contact ($3 \times 10^6/2R_{SiNP-}=16300$ g/mol/Å), like for HA/NP complexes.

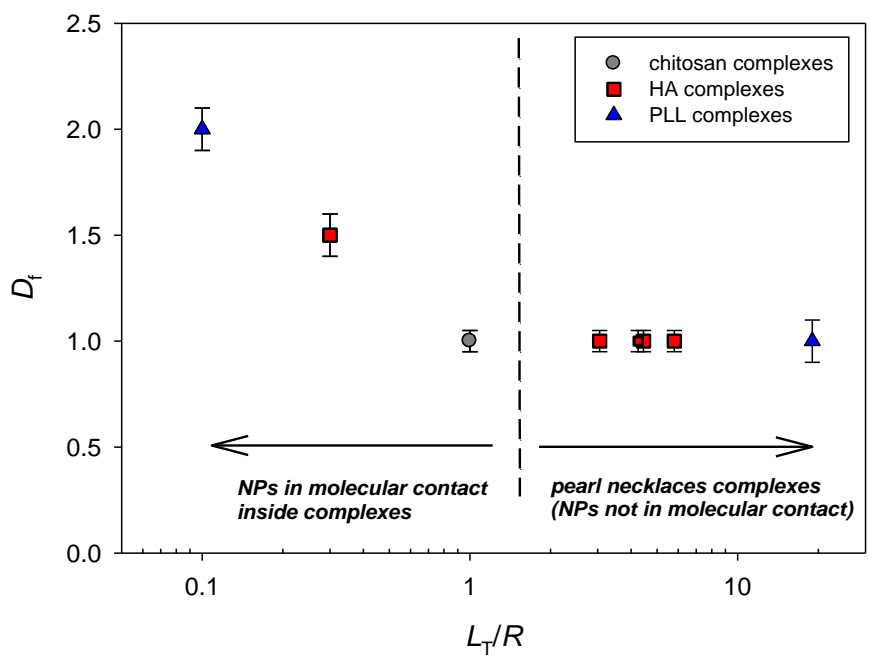

Fig. 9. (Color on-line) Influence of the ratio $L_T/R$ on the fractal exponent $D_f$ of the self-assemblies in domain III. The $L_T/R$ range is obtained by varying the nature of the PEL as well as the ionic strength $I$, which depends on the external salt, PELs, and NPs concentration. Low $I$ data are on the right hand side of the vertical dotted line. Error bars depend on both the quality of the fit of the data and of the number of samples (different runs and concentrations, see Figs. 4a, 4b, 7, and 8).

Results obtained in domains I and II (see Table 2) corroborate these observations and indicate that the complexation is weaker in any case at low $I$ and that salt-free complexes with larger $L_T/R$ adopt a looser structure. In domain I, the PEL excess may balance the NPs charge and the NPs might come in molecular contact with each other. However, the scattering signal being too low at these NPs concentrations, it is difficult to confirm this assumption with a good accuracy.

## B. BIPHASIC DOMAIN

**Biphasic domain:** Figs. 1 and 5 indicate that biphasic domains are not limited to the stoichiometry line [+]/[-]=1 (grey, blue and red lines), but on the contrary are quite wide, showing that mixtures phase separate before the neutralization point. Besides polydispersity, the reason behind this may be that primary complexes do not have the same charge ratio [+]/[-], but instead show a "disproportionation" of charge distribution [42]. This view of coexistence of neutral and more charged complexes is consistent with the observation of complexes, very likely to be still overcharged, of smaller size with same $D_f$ in some supernatants (not shown here). A wider biphasic domain, with a boundary between domains II and III located at larger NP contents, is observed in two cases:

- For the different PELs, observed first at large $I$, the widening occurs for smaller characteristic ratios $L_p/R$: for PLL, more flexible, domain II is the widest as seen in Fig. 1. Not surprisingly, this shows that the charge of NPs is more efficiently balanced by the winding of flexible PEL chains.
- For the observations on HA/NP complexes in Fig. 5, a wider biphasic part is seen for large $I$. The shaded part of domain II vanishes for low $I$: the stability of overcharged primary complexes is higher. This can be related again with the increased rigidity at low $I$, hence uneasy binding of PEL chains. But it can also be said that van der Waals attraction becomes dominant in excess of salt and causes the phase separation, in a way we now detail below.

**Effect of the interaction profile.** The origin of the difference of fractal dimension at low and high $I$ and thus of the NPs aggregation behavior can be understood by consideration of the nature of the short-range interaction energy between two approaching particles [43-45]. For charged colloids, the key feature is the repulsive electrostatic energy barrier $E_{el}$ which has to be overcome, for example, by a NP approaching a primary complex: $E_{el}$ is determined by the surface charge and the screening length fixed by the ion concentration in solution. At low $I$, complexation between oppositely charged species occurs readily and we can assume that $E_{el}$ is much less than $k_BT$. This has direct effects on the fractal dimension $D_f$: every collision will result in the particles sticking together, leading to a rapid aggregation, limited only by the rate of diffusion-induced collisions between the clusters. In this regime, called diffusion-limited colloid aggregation (DLCA), the interior of the complexes is screened from penetration, which results in a more tenuous, in other words more "open" structure with lower $D_f$. Repulsion between NPs inside complexes can also account for such "open" structures.

By contrast, if PEL-NP attraction is screened by addition of co-ions from added salt, $E_{el}$ could remain comparable to, or larger than $k_BT$, and many collisions can occur before two particles stick to one another (regime called reaction-limited colloid aggregation: RLCA). The probability of sticking $P\sim\exp(-E_{el}/k_BT)$ is much lower and the aggregating clusters will have the opportunity to explore a large number of possible mutual configurations, which leads to some interpenetration and eventually close contacts, and therefore denser NP assemblies with larger fractal exponents $D_f$. For RLCA $D_f$ is closer to 2.1, much larger than for DLCA [43-45].

In summary, we have shown the pivotal role of the quantity $L_T$ in the tuning of NP self-assemblies. The absence of external salt undoubtedly accounts for more stretched structures with lower $D_f$.

## V. CONCLUSION

To summarize, variations in $L_T/R$ allowed us to control the structure of NPs self-assemblies. Increasing electrostatic screening or reducing $L_T/R$ are two interrelated efficient ways to drive an enhancement of phase separation, through the expansion of the coacervate region, and of complex compactness. For $L_T/R<1$, fractal dimension values $D_f$ increase up to 2 in excess of NPs. $D_f=1$ for $L_T/R\geq1$, this comprises low ionic strength $I$ data but also chitosan complexes at large $I$ for which this characteristic ratio is just equal to 1. Hence, a universal account of $I$ through the total persistence length $L_T$ is obtained here. Low-dimensional NP assemblies, such as well-defined compact nanorods obtained for $L_T/R\sim1$, open routes toward new applications [1-7, 13, 14] such as plasmon-based waveguide, biosensors, nanorulers, theragnostic materials, or cancer cells therapy (e.g. with gold nanoparticles) [46]. Opposite to elaborated procedures, such as rodlike nanocrystals syntheses, this basic approach of self-assembly of preformed spherical NPs leads to nanorods with well-defined length, determined by the PEL contour length, and monodisperse cross-section (the cross-section radius and polydispersity values are similar to those of free NPs [13, 14]). On the other hand, pearl necklace nanorods obtained for $L_T/R>1$ at low $I$ could have strong interest by tuning the distance between NPs and hence some electromagnetic properties. One may thus foresee that our approach can be applied to a variety of developments involving other types of nanoparticles - such as gold and metallic ones, proteins, or viruses – covering a wide range of applications in materials as well as in the biological sciences.

**Acknowledgments.** This work was supported by a doctoral fellowship (L. S.) from the CEA and the CNRS. The authors thank the University of Paris Diderot as well as the Laboratoire Léon Brillouin and ESRF for beamtime allocations and A. Helary for his help during the SANS measurements.

[1] M. Grzelczak, J. Vermant, E. M. Furst, and L. M. Liz-Marzàn, ACS Nano **4**, 3591–3605 (2010).

[2] Z. Nie, A. Petukhova, and E. Kumacheva, Nat. Nanotechnol. **5**, 15 (2010).

[3] S. C. Glotzer and M. J. Solomon, Nat. Mater. **6**, 557 (2007).

[4] G. A. DeVries, M. Brunnbauer, Y. Hu, A. M. Jackson, B. Long, B. T. Neltner, O. Uzun, B. H. Wunsch, and F. Stellacci, Science **315**, 358 (2007).

[5] W. U. Huynh, J. J. Dittmer, and A. P. Alivisatos, Science **295**, 2425 (2002).

[6] Z. Tang, N. A. Kotov, and M. Giersing, Science **297**, 237 (2002).

[7] A. Courty, A. Mermet, P. A. Albouy, E. Duval, and M. P. Pileni, Nat. Mater. **4**, 395 (2005).

[8] N. Jouault, E. Moulin, N. Giuseppone, and E. Buhler, Phys. Rev. Lett. **115**, 085501 (2015).

[9] K. Liu, Z. Nie, N. Zhao, W. Li, M. Rubinstein, and E. Kumacheva, Science **329**, 197–200 (2010).

[10] Z. Nie, D. Fava, M. A. Winnik, M. Rubinstein, and E. Kumacheva, J. Am. Chem. Soc. **130**, 3683 (2008).

[11] S. Sacanna, W. T. M. Irvine, P. M. Chaikin, and D. J. Pine, Nature **464**, 575-578 (2010).

[12] D. Zerrouki, J. Baudry, D. Pine, P. Chaikin, and J. Bibette, Nature **455**, 380 (2008).

[13] L. Shi, F. Carn, F. Boué, G. Mosser, and E. Buhler, ACS Macro Lett. **1**, 857–861 (2012).

[14] L. Shi, F. Carn, F. Boué, G. Mosser, and E. Buhler, Soft Matter **9**, 5004-5015 (2013).

[15] L. Shi, E. Buhler, F. Boué, and F. Carn, Langmuir **31**, 5731-5737 (2015).

[16] T. C. Deivaraj, N. L. Lala, and J. Y. Lee, J. Colloid Interface Sci. **289**, 402 (2005).

[17] M. H. Huang, A. Choudrey, and P. Yang, Chem. Commun. **12**, 1063 (2000).

[18] C. Holm, J. F. Joanny, K. Kremer, R. R. Netz, P. Reineker, T. Vilgis, and R. Winkler, Adv. Polym. Sci. **166**, 67 (2004).

[19] R. R. Netz and J. F. Joanny, Macromolecules **32**, 9026 (1999).

[20] S. Ulrich, A. Lagueric, and S. Stoll, Macromolecules **38**, 8939 (2005).

[21] A. Y. Grosberg, T. T. Nguyen, and B. I. Shklovskii, Rev. Mod. Phys. **74**, 329 (2002).

[22] T. T. Nguyen and B. I. Shklovskii, J. Chem. Phys. **114**, 5905 (2001).

[23] C. Y. Kong and M. Muthukumar, J. Chem. Phys. **109**, 1522–1527 (1998).

[24] M. Jonsson and P. Linse, J. Chem. Phys. **115**, 10975 (2001).

[25] M. Jonsson and P. Linse, J. Chem. Phys. **115**, 3406 (2001).

[26] K. Keren, Y. Soen, G. Ben Yoseph, R. Gilad, E. Braun, U. Sivan, and Y. Talmon, Phys. Rev. Lett. **89**, 088103 (2002).

[27] A. A. Zinchenko, K. Yoshikawa, and D. Baigl, Phys. Rev. Lett. **95**, 228101 (2005).

[28] A. B. Kayitmazer, D. Shaw, and P. L. Dubin, Macromolecules 3**8**, 5198 (2005).

[29] I. Morfin, E. Buhler, F. Cousin, I. Grillo, and F. Boué, Biomacromolecules **12**, 859 (2011).

[30] J. Gummel, F. Cousin, and F. Boué, J. Am. Chem. Soc. **129**, 5806–5807 (2007).

[31] F. Cousin, J. Gummel, D. Ung, and F. Boué, Langmuir **21**, 9675-9688 (2005).

[32] E. Buhler and M. Rinaudo, Macromolecules **33**, 2098–2106 (2000).

[33] F. Bonnet, R. Schweins, F. Boué, and E. Buhler, Europhys. Lett. **83**, 48002 (2008).

[34] E. Buhler and F. Boué, Eur. Phys. J. E **10**, 89 (2003).

[35] E. Buhler and F. Boué, Macromolecules **37**, 1600–1610 (2004).

[36] G. M. Lindquist and R. A. Stratton, J. Colloid Interf. Sci. **55**, 45-59 (1976).

[37] E. Buhler, J. Appell, and G. Porte, J. Phys. Chem. B **110**, 6415–6422 (2006).

[38] For low NPs contents (1 or 0.5 g/l), PLL biphasic samples do not always exhibit a clear interface between the two phases.

[39] For large amounts of NPs (e.g. HA complexes with 10 g/l of SiNPs), the separation can be quick at low $I$. For lower NPs contents (e.g. 4g/l), supernatants are slightly turbid and the scattered intensity variation is a power law similar to that measured for coacervates, suggesting that the dense objects are formed before macroscopic phase separation. Complexes

remaining in the supernatants may have an overall smaller size, even if their $D_\text{f}$ is the same than that determined for coacervates.

[40] T. Odijk, J. Polym. Sci., Polym. Phys. **15**, 477 (1977).

[41] J. Skolnick and M. Fixman, Macromolecules **10**, 944 (1977).

[42] R. Zhang and B. I. Shklovskii, Physica A **352**, 216 (2005).

[43] D. A. Weitz, J. S. Huang, M. Y. Lin, and J. Sung, Phys. Rev. Lett. **54**, 1416 (1985).

[44] M. Y. Lin, H. M. Lindsay, D. A. Weitz, R. C. Ball, R. Klein, and P. Meakin, Phys. Rev. A **41**, 2005 (1990).

[45] M. Y. Lin, H. M. Lindsay, D. A. Weitz, R. Klein, R. C. Ball, and P. Meakin, J. Phys.: Condens. Matter **2**, 3093 (1990).

[46] X. Huang, M. A. El-Sayed, J. Adv. Research **1**, 13 (2010).

APPENDIX: LIGHT AND SMALL-ANGLE X-RAY SCATTERING (SAXS) CHARACTERIZATION OF THE SILICA NANOPARTICLES AND POLYELECTROLYTES

**Static and dynamic light scattering:** Fluctuations in the scattered intensity with time $I(q,t)$ (also called count rate), measured at a given scattering angle $\theta$ or equivalently at a given scattering wave vector $q=(4\pi n/\lambda)\sin(\theta/2)$, are directly reflecting the so-called Brownian motion of the scattering particles. In dynamic light scattering (DLS), the fluctuation pattern is translated into the normalized time autocorrelation function of the scattered intensity, $g^{(2)}(q,t)$. It is related to the so-called dynamic structure factor (or concentration fluctuations autocorrelation function), $g^{(1)}(q,t) = \frac{\langle \delta c(q,0) \delta c(q,t) \rangle}{\langle \delta c(q,0)^2 \rangle}$, where $\delta c(q,t)$ and $\delta c(q,0)$ represent fluctuations of the concentration at time t and zero, respectively. In the case of a diffusive process, with characteristic relaxation time $\tau$ inversely proportioned to $q^2$, $g^{(1)}(q,t)=\exp(-Dq^2 t)$, where $D$ is diffusion coefficient. The Stokes-Einstein relation allows one to determine the hydrodynamic radius $R_H$ of the scattered objects; $R_H=kT/6\pi\eta D$, if the temperature T and solvent viscosity $\eta$ are known (here $\eta=0.89$ cP at 25 °C). To determine the polydispersity index of particles, we have adopted the classical cumulant analysis: $\ln g^{(1)}(q,t) = k_0 - k_1 t + \frac{k_2}{2}t^2 + ...$, where $k_1=1/<\tau>$ and $k_2/k_1^2$ represents the polydispersity index (PDI).

In static light scattering (SLS) experiments, the excess of scattered intensity is measured with respect to the solvent. The so-called excess Rayleigh ratio was deduced using a toluene sample reference for which the excess Rayleigh ratio is well-known ($R_{toluene}=1.3522\times10^{-5}$ cm$^{-1}$ at 633 nm): $R_{solute}(cm^{-1}) = \frac{I_{solution}-I_{solvent}}{I_{toluene}} \times \left(\frac{n}{n_{toluene}}\right)^2 \times R_{toluene}$. The usual equation for absolute light scattering combines the form factor $P(q)$, the structure factor $S(q)$ and the weight-average molecular weight $M_w$ of the scattered objects: $R(q) = \frac{4\pi^2 n^2}{N_A \lambda^4}(\frac{dn}{dc})^2 CM_w P(q)S(q)$, where $K=4\pi^2 n^2 (dn/dC)^2/N_A\lambda^4$ is the scattering constant (refractive index $n=1.34$ at 25 °C for water), $C$ the concentration in g/cm$^3$, $dn/dC$ is the measured refractive index increment (obtained using a Mettler Toledo Portable Lab refractometer) and $N_A$ the Avogadro's number. The data obtained at low $q$ using SLS (see Table 1) and corresponding to large spatial scales can be fitted by a Zimm law as:

$$\frac{KC}{R(q,C)} = \frac{1}{M_W}\left(1 + q^2\frac{R_G^2}{3}\right) + 2A_2 C, \quad (A1)$$

where $R_G$ is the radius of gyration of the particles and $A_2$ the second Virial coefficient. For example, $A_2 = (2.5 \pm 0.3) \times 10^{-4}$ cm$^3$g$^{-2}$mol for 0.3 M CH$_3$COOH / 0.2 M CH$_3$COONa aqueous solutions of chitosan.

**Small-angle X-ray scattering (SAXS):** SAXS experiments were performed at the European Synchrotron Radiation Facility (ESRF, Grenoble, France) on the ID-02 beamline using a pinhole camera and 1 mm capillaries. For negatively charged SiNPs (Ludox AM), two sets of sample-to-detector distances ($D$=1 m and 8 m) were chosen at an energy of 12.46 keV corresponding to a wide $q$-range varying between 0.0011 and 0.57 Å$^{-1}$. In a second run, corresponding to the characterization of the positively charged SiNPs (Ludox CL), two configurations were chosen ($\lambda$=1 Å, $D$=1 m; and $\lambda$=1 Å, $D$=10 m) so that the following $q$-ranges were respectively available: $0.011 \leq q$ (Å$^{-1}$) $\leq 0.625$; and $9\times10^{-4} \leq q$ (Å$^{-1}$)$\leq 0.063$. Finally, a Bonse-Hart configuration allowing for a $q$-range extending from $1.32\times10^{-4}$ to 0.020 Å$^{-1}$ was used for the most concentrated samples. The absolute units are obtained by normalization with respect to water (high $q$-range) or lupolen (low $q$-range). For SAXS, the scattering length densities (SLDs) are defined by $\rho = 1/(mv\times1.66\times10^{-24})\times r_{el}\times\Sigma n_i Z_i$, where $r_{el}$=0.28×10$^{-5}$ nm is the electron radius and $Z_i$ the atomic number of element i. Table A1 reports the scattering length densities SLDs per unit volume of polyelectrolytes and silica nanoparticles calculated for SANS and SAXS.

Table A1. Scattering length densities SLDs per unit volume calculated for SANS and SAXS. The specific volumes of the different components are also indicated as well as light scattering constant $K$ and $dn/dC$ values.

| Components | Specific volume ($cm^3/g$) | Calculated SLD-SANS/$\times 10^{10}$ $cm^{-2}$ | Calculated SLD-SAXS/$\times 10^{10}$ $cm^{-2}$ | $dn/dC$ ($cm^3/g$) | $K$ ($cm^2 \cdot g^{-2} \cdot mol$) |
|---|---|---|---|---|---|
| $H_2O$ | 1 | -0.56 | 9.37 | - | - |
| $D_2O$ | 0.9058 | 6.36 | 9.31 | - | - |
| Chitosan in 0.3M $CH_3COOH$/0.2M $CH_3COONa$ | 0.478 | 2.47 for $C_6H_{11}NO_4$ (in $H_2O$) 5.32 for $C_6H_7NO_4D_4$ (in $D_2O$) | 18.7 | 0.195 | $2.79 \times 10^{-7}$ |
| Poly-L-lysine in 0.2M KBr | ~1 | 1.06 for $C_6H_{12}N_2O$ | 9.3 | 0.1645 | $1.98 \times 10^{-7}$ |
| Hyaluronan in 0.1M NaCl | 0.59 | 2.37 for $C_{14}H_{21}NO_{11}$ | 15.1 | 0.14 | $1.44 \times 10^{-7}$ |
| Silica NPs (Ludox AM 30: negatively charged) in 0.3M $CH_3COOH$/0.2M $CH_3COONa$ | 0.4545 | 3.47 | 18.5 | 0.0658 | $3.17 \times 10^{-8}$ |
| Silica NPs in 0.1M NaCl at pH=4 (Ludox CL: positively charged) | 0.4545 | 3.47 | 18.5 | 0.064 | $3 \times 10^{-8}$ |

**Negatively charged SiNPs:** Ludox AM particles, provided by GRACE, carry a pronounced negative surface charge over the whole pH range above the isoelectric point (located around pH=2) due to the substitution of tetravalent silicium by trivalent aluminium ions ([$Al_2O_3$]= 0.2 wt. % according to the supplier) giving rise to very good stability against pH variation [14]. Such stability has been checked using light scattering measurements showing a narrow size distribution over several months. SiNPs are well dispersed and stable in solution and show no tendency to aggregation with time. We characterized the scattering from a SiNPs dilute suspension, introducing a polydispersity in size of the scattered objects described by a log-normal distribution, $L(r, R, \sigma)$, where $r$ is the radius, $R$ the mean radius, and $\sigma$ the variance:

$$L(r,R,\sigma) = \frac{1}{\sqrt{2\pi}r\sigma}\exp\left(-\frac{1}{2\sigma^2}\ln^2\left(\frac{r}{R}\right)\right) \quad (A2)$$

Thus, neglecting the virial effects (assuming $S(q) = 1$) at low concentration in the presence of salt, it is classical to define the global scattering intensity by the following relation:

$$I(q) = \phi(\Delta\rho)^2 V \int_0^\infty P(q,r)L(r,R,\sigma)dr \quad (A3)$$

Fig. A1 shows the scattering of a pure SiNPs solution, which can be fitted satisfactorily by means of the form factor expression derived for hard spheres of radius $R$:

$$P(q) = 9 \times \left[\frac{\sin(qR) - qR\cos(qR)}{(qR)^3}\right]^2 \quad (A4)$$

The form factor oscillations, damped by the size distribution, are well-reproduced with $I(q)$ calculated as indicated above (eqs A3 and A4). The negatively charged SiNPs solution is well-represented by a suspension of hard spheres with $R=9.2$ nm and $\sigma=0.12$. Extrapolation of the scattered intensity to zero-wave vector, $I(0)$, gives the SiNPs weight-average molecular weight, $M_W=3\times10^6$ g·mol$^{-1}$.

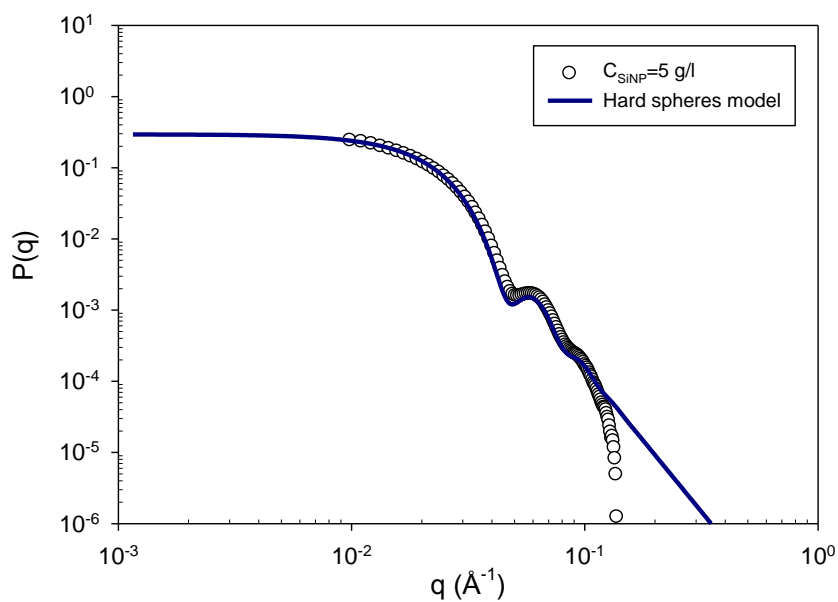

Fig. A1. (Color on-line) SiNP form factor, $P(q)$, obtained using SAXS experiments performed at 5 g/l and 20 °C. The continuous blue line corresponds to the best fit of the data using the form factor expression derived for hard spheres of radius $R$=9.2 nm with $\sigma$=0.12.

**Positively charged SiNPs:** Commercial dispersion of positively charged SiNPs was provided by GRACE under the reference Ludox CL-CAL25 and was used as received without further purification. Ludox CL is a colloidal dispersion of modified silica particles coated with aluminum oxide $Al_2O_3$ (3% according to the supplier) using aluminum chloride to reverse the surface charge density. Therefore, these modified silica particles carry a more pronounced positive charge on their surface (Zeta potential is equal to 17.8 mV in the presence of 0.1 M NaCl). Positive $Al^+$ ions are bonded to the SiNPs surface through the silanol groups. We have considered a NP surface charge density of 1 elementary charge per $nm^2$. The original 30 wt.% stock was diluted into a $10^{-4}$ M HCl solution to ensure a pH value of 4 and thus a good stability of the particles at high (0.1 M NaCl) and low (salt-free solution) ionic strength. Analysis of the single cooperative relaxation mechanism measured by DLS gives an apparent hydrodynamic radius, $R_H$, of 22±1 nm, a value independent on the excess salt concentration. These silica particles are monodisperse in size as seen by the polydispersity index calculated using the cumulant procedure: $k_2/k_1^2$=0.057. Their stability with time was also checked using DLS by performing new experiments 7 days after samples preparation, and no changes in the autocorrelation function were observed.

Fig. A2 shows SAXS profiles collected at different SiNP concentrations and at pH=4. The spectra are identical, showing that the structure of the SiNPs remains unchanged in whole investigated concentration range. Additionally, the presence of an excess of salt does not destabilize the SiNPs dispersion either as seen by the superimposition of the data obtained for a $C$=10 g/l solution prepared with and without addition of salt.

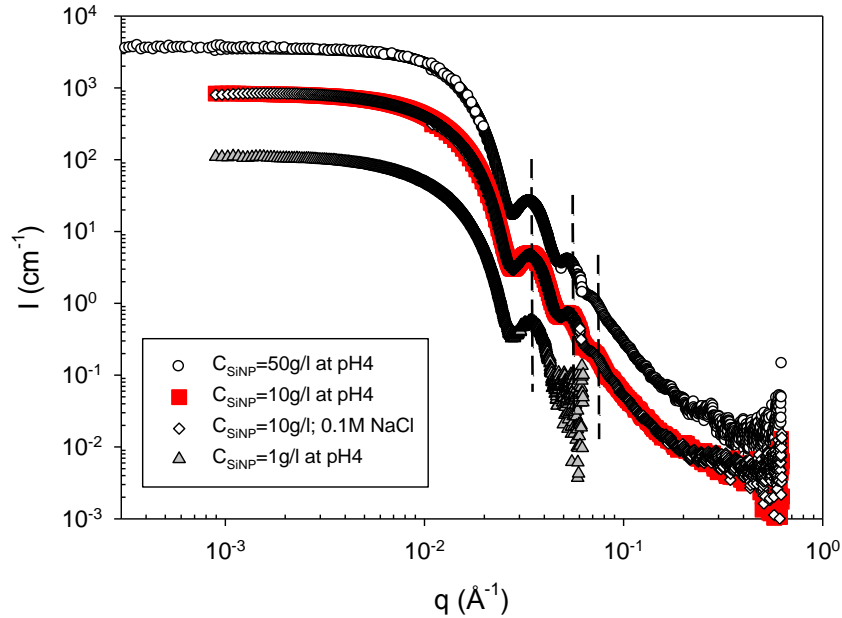

Fig. A2. (Color on-line) SAXS profiles collected at various positively charged SiNP concentrations. The dashed lines indicate the oscillations associated to the form factor of the SiNPs.

The low-q range data can fitted by a classical Guinier expression, which provides the average radius of gyration, $R_G$, equal to 16 nm for all concentrations, and the zero-wave vector scattered intensity, $I(0)$, associated to the mass of the particles. Extrapolation of the ratio $C/I(q^2 \rightarrow 0)$ to zero-concentration gives $M_W$=3.37×10$^7$ g/mol and $A_2$=4.54×10$^{-5}$ cm$^3$.g$^{-2}$.mol (see eq. A1), a positive value indicating that the pH 4 buffer is a good solvent for the cationic SiNPs. The high $q$ data can be fitted satisfactorily by means of the form factor expression derived for hard spheres of radius $R$ (see eq. A4). One obtains $R$=17 nm (data are summarized in Table 1).